\documentclass{article}
\usepackage{graphicx}
\pdfoutput=1
\usepackage[%
  pdftitle={Fundamentals of LHC Experiments},%
  pdfauthor={Jason Nielsen},%
  pdfsubject={},%
  pdfkeywords={LHC, TASI, particle physics, detectors},%
  breaklinks=true,%
  colorlinks=true,%
  linkcolor=blue,anchorcolor=blue,%
  citecolor=blue,filecolor=blue,%
  menucolor=blue,pagecolor=blue,%
  urlcolor=blue]{hyperref}

\title{\vspace{-2cm}\begin{flushright}
       \mbox{\normalsize SCIPP 11/06}
       \end{flushright}
       \vspace{1.5cm} Fundamentals of LHC Experiments}
\author{Jason Nielsen\\Santa Cruz Institute for Particle Physics and
  Department of Physics\\ University of California, Santa Cruz, CA
  95064 \\ {\tt nielsen@scipp.ucsc.edu}} 
\date{June 1, 2011}

\begin{document}
\maketitle

\begin{abstract}
Experiments on the Large Hadron Collider at CERN represent our
furthest excursion yet along the energy frontier of particle physics.
The goal of probing physical processes at the TeV energy scale puts
strict requirements on the performance of accelerator and experiment,
dictating the awe-inspiring dimensions of both.  These notes, based on
a set of five lectures given at the 2010 Theoretical Advanced Studies
Institute in Boulder, Colorado, not only review the physics considered
as part of the accelerator and experiment design, but also introduce
algorithms and tools used to interpret experimental results in terms
of theoretical models.  The search for new physics beyond the Standard
Model presents many new challenges, a few of which are addressed in
specific examples.
\end{abstract}

\section{Introduction\label{sec:intro}}

Experimental results combined with theoretical considerations imply
the existence of new physics beyond the Standard Model, at energies no
greater than 1\,TeV.  Although this has been known for a
while~\cite{Eichten:1984eu}, the possibility of accessing this energy
scale, known as the ``terascale,'' has now been realized in
current-day hadron colliders and the experiments that use them.  These
notes provide a brief overview of the experimental considerations and
design needed to measure particle interactions at the terascale.

To probe directly the physics at the 1 TeV scale, we need a momentum
transfer $Q^2$ of approximately 1 TeV between the initial state
particles.  This direct requirement assumes we are most interested in
producing real (on-shell) new particles.  There is still a role, of
course, for precision experiments on the intensity frontier that
measure the effects of off-shell new particles in loop diagrams, but
that lies outside the current discussion of the energy frontier.

\section{Proton Beams for Terascale Physics\label{sec:beams}}

To achieve the goal of beams for terascale physics, we consider three
possibilities.  First, we could pursue collisions of high-energy beams
on stationary targets, but the center-of-mass energy in such
collisions scales only as $\sqrt{E_\mathrm{beam}}$ , so this option
seems impractical.  Second, we might investigate a lepton-antilepton
collider, in which the center-of-mass energy scales with
$E_\mathrm{beam}$.  Third, we might pursue a hadron-hadron collider,
either with proton-proton or proton-antiproton collisions.  Because
the hadrons are composite particles with varying constituent momenta,
this approach is somewhat less convenient than a lepton-antilepton
collider.  Nevertheless, we shall see that other considerations favor
the hadron collider concept as implemented in the Large Hadron
Collider.

Synchrotron radiation limits the choices of lepton-antilepton
colliders, since the energy radiated by a charged particle in each
turn of a circular accelerator is proportional to $\gamma^4/\rho$.
For example, radiation losses per turn were $0.2\,\mu\mathrm{W}$ for
each electron in the LEP collider $45\,\mathrm{GeV}$ beams.  Even
though the losses can be overcome, replacing the energy through
acceleration after each turn is prohibitively inefficient.  For these
reasons we expect high-energy colliders either to accelerate more
massive charged particles (protons or muons) or to push the ring
radius $\rho$ to its ultimate limit in a linear collider.

There are two caveats relating to linear colliders.  First, a linear
collider seems to require full accumulation of beam particles and full
acceleration for each shot, unless the beam particles can be recycled.
Second, accelerating beams to 500 GeV (the energy required for a 1 TeV
lepton collider to probe the ``terascale'') requires very high
acceleration gradients over a very long linear path.  Given that the
state of the art in radiofrequency gradients is about
$30\,\mathrm{MV/m}$, accelerating a single beam to $500\,\mathrm{GeV}$
would take about $15\,\mathrm{km}$.  Work is ongoing to improve those
gradients for a high-energy linear electron-positron collider.

The rest of our discussion will focus on circular proton or antiproton
colliders.  Whereas accelerating non-relativistic particles requires
only a bending magnetic field fixed to $B=p/q\rho$ and a constant
accelerating cyclotron frequency $\omega_c = qB/m$, accelerating
relativistic particles requires an accelerating frequency
$\omega_c=qB/\gamma m$.  During acceleration, then, ramping up the $B$
field bends the beams of increasing momentum within a fixed bending
radius while allowing a fixed accelerating frequency to be used.  In
MKS units, $p=0.3 B\rho$, where the momentum $p$ has units of GeV/c,
$B$ has units of Tesla, and $\rho$ has units of meters.  There are
obvious limits to both $B$ and $\rho$, some physical and some
geographical.

If our goal is to have a proton beam of at least $3.5\,\mathrm{TeV}$ in
order to access the terascale physics at $Q^2 > 1\,\mathrm{TeV}$, we
must work within these limitations.  To maximize the bending radius,
we might choose the large underground ring at CERN (an octagon with
alternating straight accelerating and bending sections), which has a
$4.3\,\mathrm{km}$ average radius.  In this ring, a
$3.5\,\mathrm{TeV}$ beam requires an average bending field of
$4.2\,\mathrm{T}$, while a $7\,\mathrm{TeV}$ beam requires
$5\,\mathrm{T}$.  (Since the bending magnets are not distributed
everywhere around the ring, the peak requirements are somewhat higher,
up to $8.3\,\mathrm{T}$ for the $7\,\mathrm{TeV}$ case.)  These fields
are far above the domain saturation cutoff for regular ferromagnets,
but superconducting magnets can achieve close to $10\,\mathrm{T}$ in
the steady state, limited by intercable and interfilament forces.  A
heavy laboratory-industrial partnership has developed NbTi
``Rutherford cable'' needed to bend beams of the required momentum in
the CERN ring now occupied by the Large Hadron Collider.

Given the possibility of accelerating and bending high-energy beams in
a large enough ring, what beam intensities are needed for studies of
the rare interactions of the terascale?  The answer depends on several
variables, including the beam size, particle spacing, and potential
effects on the experiments.  The small interaction rates expected at
$Q^2=1\,\mathrm{TeV}$ require instantaneous luminosities of
$10^{34}\,\mathrm{cm}^{-2}\mathrm{s}^{-1}$ to collect enough events to
study.  With the design parameters of the LHC, this corresponds to
$3\times 10^{14}$ protons per beam.  (These required luminosities are
much greater than those available at the Fermilab Tevatron, where the
proton-antiproton collisions have instantaneous luminosities of
$4\times 10^{32}\,\mathrm{cm}^{-2}\mathrm{s}^{-1}$.)

Up to this point we have kept open the possibility of
proton-antiproton or proton-proton collisions, but now we have to make
a choice.  Antiproton-proton collisions have some advantage, as they
enhance the $q\bar q$ initial states due to the dominant parton
distributions at high parton momentum.  Unfortunately it seems
unfeasible to pack $3\times 10^{14}$ antiprotons into a single beam.
At the Tevatron, 25 years of experience producing and accumulating
antiprotons has resulted in a maximum accumulation rate of $3\times
10^{10}$ antiprotons per hour, giving a maximum antiproton count of
$3\times 10^{12}$ in a beam.  This is a factor of 100 too low of the
value needed to produce events in the rarest interactions.  As a
result, to investigate rare processes with $Q^2>1\,\mathrm{TeV}$
occurring at low rates, the proton-proton collider is the only choice
at present.

\begin{figure}
  \centerline{\includegraphics[scale=0.4]{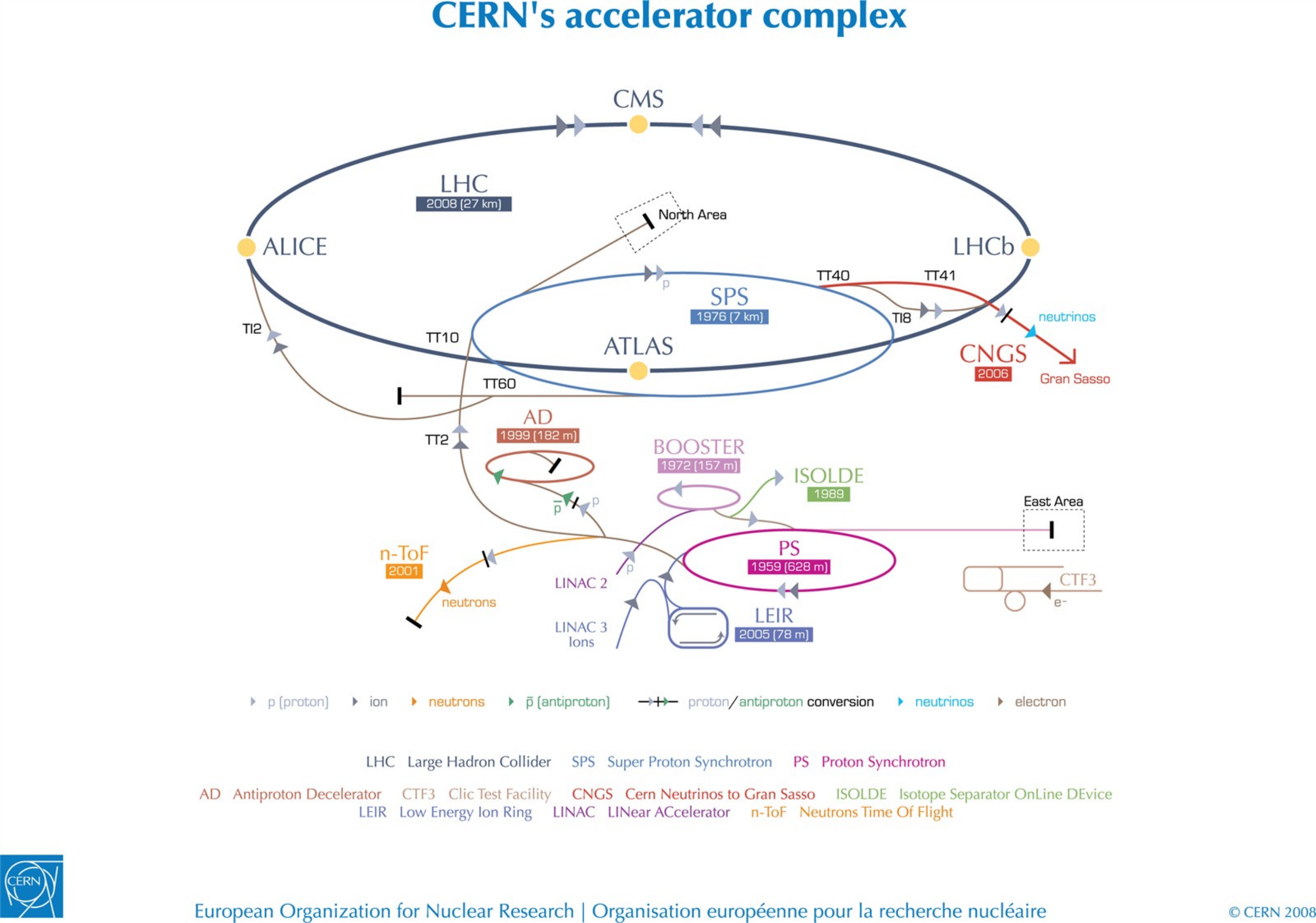}}
  \caption{Overview of the CERN accelerator chain, showing the
    relationship between low-energy accelerators and the Large Hadron
    Collider with experiments.  Image credit: CERN\label{fig:cernaccel}}
\end{figure}

The Large Hadron Collider is the final stage of the CERN accelerator
chain shown in Fig.~\ref{fig:cernaccel}.  Protons are taken from a
single bottle of hydrogen gas, accelerated in the linac and Proton
Synchrotron to $26\,\mathrm{GeV}$, and injected into the Super Proton
Synchrotron.  After being accelerated in the SPS to
$450\,\mathrm{GeV}$, protons are injected into the twin rings of the
LHC.  The staged injection energies for each accelerator minimize the
required operational range for the bending magnets.

The LHC itself is composed of two rings, one for each proton beam
direction.  Because the existing CERN tunnel has a diameter of just
$3.7\,\mathrm{m}$, a ``twin-bore'' design first proposed by Blewett is
used.  In this design both rings are contained in a
single superconducting cold mass and cryostat, but the magnetic dipole
fields point in opposite directions to provide Lorentz bending force
toward the center of the ring.  An important part of the magnet design
is the total dipole length of $15\,\mathrm{m}$, chosen to reduce the
number of inter-cryostat connections.

The high magnetic field and large volume of these dipoles imply an
enormous amount of stored magnetic energy in each magnet.  A simple
calculation per magnet of
\begin{equation}
E=\frac{1}{2} LI^2 = \frac{1}{2}
(0.099\,\mathrm{H})(11.8\,\mathrm{kA})^2 = 6.9\,\mathrm{MJ}
\end{equation}
demonstrates the need for a quench protection system.  The
superconducting cable quenches and becomes a regular ohmic conductor
if the temperature or magnetic field rise above critical values.  If
the current is not carried away through shunt resistances, it has the
potential to boil off liquid helium explosively.

The protons beams themselves are accelerated with radio frequencies of
400 MHz, giving rise to synchroton oscillations that group protons
into RF ``buckets.''  The collision rates in the interaction regions
are proportional to $N_\mathrm{p}^2 n_b$, where $N_\mathrm{p}$ is the
number of protons per bunch and $n_b$ is the number of bunches.  The
LHC design goal is 2808 bunches.  Within the beams, the proton
population is distributed in both position space and momentum space,
and the ``emittance'' is a measure of that spread.  The longitudinal
emittance relates to the bunch length, while the transverse emittance
relates to the bunch (beam) width.  The LHC beams are squeezed as they
approach the interaction points, with the ``strength'' of the
squeezing gradient given by $\beta^\star$.  Formally, $\beta^\star$ is
the distance from the interaction point where the beam width doubles,
and a lower value of $\beta^\star$ means the beam is smaller at the
interaction point.  During early LHC operations a typical value of
$\beta^\star$ has been $1.5\,\mathrm{m}$.

To calculate the expected rate of collisions in the LHC, we define the
instantaneous luminosity $L$, measured in units of
area$^{-1}$time$^{-1}$.  Ultimately phase 1 of the LHC will reach
instantaneous luminosities of
$10^{34}\,\mathrm{cm}^{-2}\mathrm{s}^{-1}$ at full intensity.  We can
also write $L$ in units of cross section, using the definition
$1\,\mathrm{barn}=10^{-24}\,\mathrm{cm}^2$ or
$1\,\mathrm{nb}=10^{-33}\,\mathrm{cm}^2$, to estimate the rate of
specific physics interactions.  For example, with
$L=10^{33}\,\mathrm{cm}^{-2}\mathrm{s}^{-1}=1\,\mathrm{nb}^{-1}\mathrm{s}^{-1}$
and inelastic proton-proton interaction cross section
$\sigma=70\,\mathrm{mb}$ we expect a rate of $L\sigma=70\times
10^{6}$ interactions per second.  Since the experiments collect data
over an extended period of time, the integrated luminosity
\begin{equation}
\mathcal{L}=\int L dt
\end{equation}
is defined with units $\mathrm{area}^{-1}$, and then the dimensionless
product $\mathcal{L}\sigma$ is the total number of
interactions.

\begin{quote}
{\em Exercise 1:} Compare the center-of-mass energy in
electron-positron collisions for (a) head-on collision of particles,
each with energy $E_{\rm beam}$, and (b) collision of a beam particle
with energy $E_{\rm beam}$ on a fixed target particle of mass $m_e$.
\end{quote}

\begin{quote}
{\em Exercise 2:} Starting from the classical formula for the radiated
power from an accelerated electron, show that the loss per turn due to
synchrotron radiation is $\Delta E= 4\pi e^2 \gamma^4/3\rho$, where
$\rho$ is the ring radius and $\gamma$ is the Lorentz factor of the electron.
\end{quote}

\section{Particle Detectors for the Energy Frontier\label{sec:detectors}}

With the necessary beams designed and constructed to produce rare
interactions sensitive to TeV-scale physics, we turn to the challenge
of measuring the high-energy particles in the final states of those
interactions.  It is especially instructive to consider how particle
detectors work from the point of view of the fundamental interactions.
This allows us to predict how existing detectors would respond to new
unusual particles in theories beyond the Standard Model.

In a nutshell, the ultimate goal of particle detection is to measure
the 4-momentum of each final state particle.  This can be accomplished
by measuring $(p_x,p_y,p_z,m)$, $(p_x,p_y,p_z,E)$, or
$(p_T,\eta,\phi,E)$.  This last form makes use of the definitions for
transverse momentum $p_T^2=p_x^2+p_y^2$ and pseudorapidity
$\eta=-\ln\left[\tan(\theta/2)\right]$.  The relativistic momentum can
be measured in a magnetic spectrometer, while the velocity in the lab
frame can be measured in certain cases with precision timing
circuits.  Particle mass can be inferred from the energy loss through
ionization, and the scalar energy can be measured in a shower of large
cross section interactions.

It is clear that there must be {\em some} interaction of the final
state particles with the detector medium, and we make use of the two
highest-rate interactions -- the electromagnetic and strong
interactions.  These two interactions affect the particles on two very
different energy scales, since the electromagnetic interaction occurs
on the atomic length scale (eV energy scale) and the strong interaction
occurs on the nuclear length scale (GeV energy
scale)~\cite{Green:2000}.

A magnetic spectrometer is at the heart of every LHC experiment.  The
Lorentz force causes charged particles to move in helical trajectories
in a solenoidal field.  Measuring the sagitta of the projected helix
gives a clean estimate of the particle's transverse momentum, using
the relation
\begin{equation}
s = \frac{0.3}{8} \frac{L^2 B}{p_T}
\label{eqn:sagitta}
\end{equation}
where $L$ is the total arc length over which the sagitta is measured.
In general, the uncertainty on the transverse momentum scales as
$\sigma_{pT}/p_T\sim p_T$.  Requirements on momentum resolution,
dictated by physics goals, translate directly to requirements on
sagitta resolutions.  As an example, the ATLAS collaboration set a
goal of 10\% momentum resolution for $1\,\mathrm{TeV}$ muons expected
from some new physics signatures.  This implies a resolution of
$50\,\mu\mathrm{m}$ on the sagitta measurement for trajectories that
are nearly straight in the muon spectrometer.  Increasing the magnetic
field $B$ or the arc length $L$ (which is effectively the detector
radius for straight tracks) increases the sagitta and leads to smaller
relative uncertainties.  This motivates large detectors like the ATLAS
muon spectrometer, which has a maximum radius of $20\,\mathrm{m}$.

The interaction underlying most non-destructive measurements is the
electromagnetic interaction, whether a particle scatters elastically
off charges in material or loses energy through ionization of atomic
electrons.  In the former case, scattering from multiple charge
centers results in an uncertainty on the particle's original momentum
vector.  This uncertainty scales with the square root of the number of
scatterers -- $\left< \theta_\mathrm{MS}^2\right> = N \left<
\theta^2\right>$, where $\theta$ is the deflection expected from a
single scatter -- and it is a powerful motivation for limiting the
amount of material in the particles' path.  In the latter case, the
energy loss follows the Bethe-Bloch formula, which in one form for a
singly-charged particle looks like
\begin{equation}
\frac{dE}{d\rho x} \sim \left(\frac{N_\mathrm{Avo} Z}{A}\right)
\left(\frac{\alpha^2 \hbar}{m_e c}\right) \left( \frac{1}{\beta^2}
\right)
\end{equation}
This energy loss translates directly to a number of ionization
electrons along the particle trajectory.

Charged particle detectors used controlled electric fields to collect
ionization electrons and cations.  In some cases, most notably in
detectors with gaseous media, a central sense wire lies at the center
of a radial electric field.  As ionization electrons drift toward the
center, the strongest part of the electric field, they are accelerated
and induce an avalanche of additional ionization.  This multiplication
factor makes it possible to observe the passage of a particle in a
relatively low-density medium.  The position resolution of such
detectors depends on exact knowledge of the drift time and is limited
by placement accuracy of the central wires.  Another detector type
uses solid-state semiconductors, usually silicon, as the interaction
medium.  Particles passing through the semiconductor ionize valence
electrons and create electron-hole pairs, which drift under an applied
electric field to strip or pixel readout elements patterned on the
surface.  These detectors benefit from increased ionization energy
loss (roughly $400\,\mathrm{keV/mm}$), but there is no amplification
of collected charge, and special low-noise readout electronics are
required to detect the signal.  A major advantage of these detectors
is that the readout element patterns can be microns apart, yielding
extremely good position resolution.  The disadvantages with respect to
gaseous detectors are cost (gas is much cheaper than lithographed
silicon) and processing requirements.  Any defects in the
semiconductor crystal trap the charge carriers and prevent them from
reaching the readout elements.  Because of their high cost and
precision position resolution, solid-state detectors are most often
placed near the interaction point, where they measure the first points
of the particle trajectories before any scattering can take place.

By measuring the energy of a particle and either $\gamma$ or $\beta$,
we can use the relation $E=\gamma m$ to determine the particle mass
and therefore the particle type.  Measurements of $\gamma$ and $\beta$
come from energy loss ($dE/dx$) as collected by readout electronics,
direct velocity measurements via precision timing, or radiation
emitted by charged particles in a dielectric medium (Cherenkov or
transition radiation).

Energy measurements in calorimeters are by nature destructive
measurements, since their aim is to fully contain and collect the
energy of the incoming particle.  There are two different mechanisms
by which a particle entering the dense material of the calorimeter
showers into a large number of second particles.  The first mechanism,
the electromagnetic shower, proceeds by alternating electron
bremsstrahlung to photons and subsequent conversion to
electron-positron pairs.  Only photons and electrons participate in
the electromagnetic shower, since both bremsstrahlung and pair
production are maximized for low mass particles.  The second
mechanism, the hadronic shower, is due to nuclear interactions.
For both mechanisms, the cascade undergoes exponential branching until
the energy of individual particles falls below some critical energy
and ionization takes over.

Because of the exponential growth of the shower, the total number of
secondary particles is proportional to the energy of the primary
particle that entered the calorimeter.  The energy resolution of the
calorimeter, then, is given by
\begin{equation}
\frac{\sigma_E}{E} \sim \frac{\sigma_N}{N} \sim \frac{1}{\sqrt{E}}
\end{equation}
since the fluctuations in the large number of particles follow a
Gaussian distribution.  The secondary particles in the shower can be
collected in a dedicated active medium distinct from the absorber
material (as in a non-homogeneous calorimeter) or in the same medium
that serves as absorber (in a homogeneous calorimeter).  This
distinction has a substantial effect on the energy resolution; for
example, the electromagnetic energy resolution is $\sigma_E/E =
10\%/\sqrt{E}$ in ATLAS but $2.7\%/\sqrt{E}$ in CMS.  In the LHC
experiments, two distinct calorimeters are deployed, one representing a small
number of nuclear interaction lengths but large number of
electromagnetic radiation lengths, and one with a large number of
nuclear interaction lengths.  These are the electromagnetic and
hadronic calorimeters, respectively.

\begin{figure}
  \centerline{\includegraphics[scale=0.35]{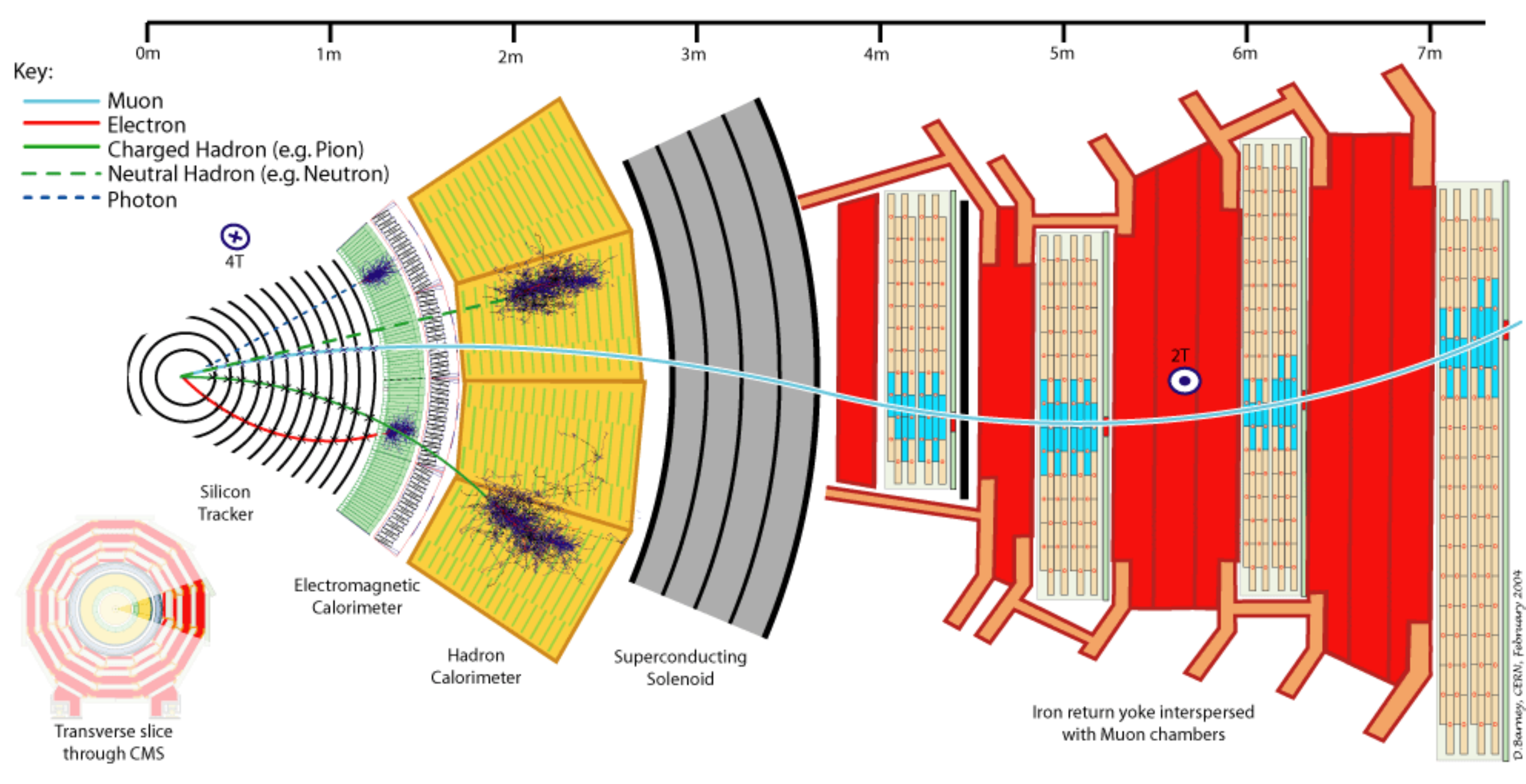}}
  \caption{Transverse slice through the CMS detector, showing the
    individual detector subsystems and particle signatures in each.
    The particle type can be inferred by combining the detector
    response in the different subdetectors.  Image credit:
    CERN\label{fig:cmsid}}
\end{figure}

Particles that do not interact strongly, such as muons and neutrinos,
penetrate the calorimeters without showering.  Muons are detected in
standalone external spectrometers or chambers integrated in the magnet
yoke.  Neutrinos do not interact with any of the detector material,
and any missing momentum in the collision is attributed to them.
Figure~\ref{fig:cmsid} shows in detail the experimental signature of
the Standard Model particles, as detected in the CMS experiment.

The high interaction rates required to search for new physics at the
TeV scale present extra challenges for the LHC experiments.  Charged
particle tracking algorithms are designed to function with detector
occupancies of up to $\mathcal{O}(1\%)$.  These algorithms, which work
by stringing together significant energy deposits (``hits''), start
with the highest-granularity silicon detectors near the interaction
point and work outward, accounting for energy loss in each detector
layer encountered.  The experiments have been designed to meet the
requirements of low occupancy even in particle-dense environments like
boosted jets from high-mass resonances or Higgs boson decays.

Certain examples of physics beyond the Standard model give rise to
striking experimental signatures, and it is worth looking at a few
such examples.  First, exotic charged massive particles
(CHAMPs) typically have low velocities, with $\beta$ significantly less
than 1.  This causes them to lose greater amounts of energy through
ionization, and they may even stop in the middle of the detector if
the energy loss is great enough.  Since the particles are massive they
do not shower in the electromagnetic calorimeter.  R-hadrons, stable
particles with heavy colored constituents, can have similar
signatures.  Doubly-charged particles lose $q^2=4$ times the normal
ionization energy loss in the tracking detectors, a unique signature
for any experiment with $dE/dx$ sensitivity.  Second, metastable
particles, such as long-lived neutralinos in theories of supersymmetry
breaking, may have $c\tau$ values of order $1\,\mathrm{m}$, and the
decay products do not point back to the primary interaction vertex.
Even though it is challenging to reconstruct particles originating in
the middle of the detector, successful reconstruction makes it
possible to measure the lifetime of the parent particle.  Third,
exotic ``quirks,'' predicted in new theories of strong interactions,
can appear as a mesoscopic bound state, with properties similar to a
doubly-charged particle, or as a state that oscillates like a
macroscopic string with two charge endpoint particles.

These and other particle interactions can be modeled with dedicated
computer programs of differing complexity.  On one end lies
GEANT4~\cite{Agostinelli:2002hh,Allison:2006ve}, a simulation toolkit
with detailed lists of high- and low-energy interactions.  Each
particle is tracked step-by-step through a custom model of the
detector.  On the other end lie fast simulation programs like
PGS~\cite{Conway:2009} and Delphes~\cite{Ovyn:2009tx}, which use
parameterized detector response to each type of particle.

Careful consideration of the physical interactions behind the
experimental techniques allows one to extrapolate detector behavior to
the most unusual new possibilities!  Unfortunately experiments do not
have large enough computing budgets to store the detector response to
each bunch crossing, so they use a multi-level trigger system to
decide immediately which events should be saved.  This bears
repeating: unless the new physics is selected by a trigger algorithm,
it will be lost forever.  

The LHC experiments use a combination of low-level hardware and
high-level software triggers to filter events for further study.  In
the hardware triggers, individual trigger objects (jets,
electromagnetic clusters, muons) are identified using data from fast
readout detectors.  These are passed to streamlined versions of
offline software algorithms to be reconstructed more fully.  For
example, an electron trigger might require a Level-1 electromagnetic
cluster with $E_T>20\,\mathrm{GeV}$, followed in the High-Level
Trigger with a set of track-matching requirements and further cluster
shape cuts to reject jets and $\pi^0$ mesons.  If the rate of electron
events becomes too great, then some filter has to be tightened to
reduce the overall rate.

\begin{quote}
{\em Exercise 3:} Beginning from the Lorentz force law, prove the
relation in Eqn.~\ref{eqn:sagitta} between the sagitta and transverse
momentum of a charged particle.
\end{quote}

\begin{quote}
{\em Exercise 4:} In silicon strip detector with strip pitch (spacing)
$d$, a hit on a strip means a particle passed somewhere in the
$\{-d/2,+d/2\}$ range centered on the strip.  Show that the variance
of this uniform distribution is the same as that of a Gaussian
distribution with width $d/\sqrt{12}$.
\end{quote}

\section{Physics Studies with Hadronic Jets\label{sec:jets}}

Many Standard Model and new physics signatures include hadronic jets.
These jets present many challenges for the LHC experiments, and since
they are the result of peturbative and non-perturbative QCD effects
they also test our theoretical framework.  Because the jets are
measured chiefly in the calorimeters, experimentalists develop special
energy calibrations to account for the effects of hadronization and
contributions from pileup.

Even though we do not expect a one-to-one correspondence between
reconstructed jets and partons (except possibly for the highest energy
partons) we do expect that the relation between the two should be
calculable on average.  In the past decade experimentalists have
become more sophisticated in defining jets and comparing measurements
with theoretical predictions.  Such comparisons are always made at the
``particle-level'' or ``hadron-level,'' after the parton shower and
hadronization but before the detector interaction occurs.  This means
that theoretical predictions must apply a showering and hadronization
model to parton-level results, and experiments must unfold the effects
of jet reconstruction in the detectors.

We have first to define exactly what is meant by ``jets,'' and there
are three main considerations.  First, what particles should be
clustered, or what inputs will be given to the algorithms?  Second,
which particles should be combined into each jet, based on proximity
and energy?  Third, how should the input 4-momenta be combined?  The
goal is to define jet clustering algorithms that are fast, robust
under particle boosts, and able to deal with collinear and infrared
radiation.  This problem has been studied in great
detail~\cite{Blazey:2000qt}, but it is useful to summarize some of the
solutions.

The answer to the first question is that the inputs can in fact be any
kind of 4-momentum, possibly associated with truth particles (to give
truth jets) or calorimeter cells (reconstructed jets).  The answer to
the third question seems to have been decided in favor of adding
4-momenta vectorially.  There are many answers to the second question
of how to choose the particles to combine into each jet!

Perhaps the simplest way to cluster particles is to use a cone
algorithm, in which a distance $\Delta=\sqrt{(\Delta \eta)^2 +
  (\Delta\phi)^2}$ is defined with respect to a seed particle
(typically the highest-$p_T$ particle).  All particles satisfying
$d<R$ lie in a circle of radius $R$ in the $\eta-\phi$ plane of the
calorimeter, and these particles are combined to form a jet
4-momentum.  This algorithm is well-defined geometrically, but the
choice of seed particle is not stable when collinear radiation is
considered.

A more sophisticated class of algorithms combine nearest particles
first, effectively reversing the branching of the parton shower,
instead of fixing a seed particle and cone.  These ``sequential
recombination'' algorithms cluster particles with smallest $d_{ij}$
first, where 
\begin{equation}
d_{ij} = \min\left(k_{ti}^{2p}, k_{tj}^{2p} \right) \frac{\Delta_{ij}^2}{R^2}
\end{equation}
for particles $i,j$ and beam 4-vector $t$.  (Particles whose closest
neighbor is the beam are considered stable jets.)  In this general
formulation, different values of the exponent $p$ give very different
algorithms.  For $p=0$ (Cambridge-Aachen algorithm), particles near
each other in $\eta,\phi$ coordinate space are clustered first,
whereas for $p=1$ (kT algorithm) lower-momentum particles are
clustered first.  

\begin{figure}
  \centerline{
    \includegraphics[scale=0.2]{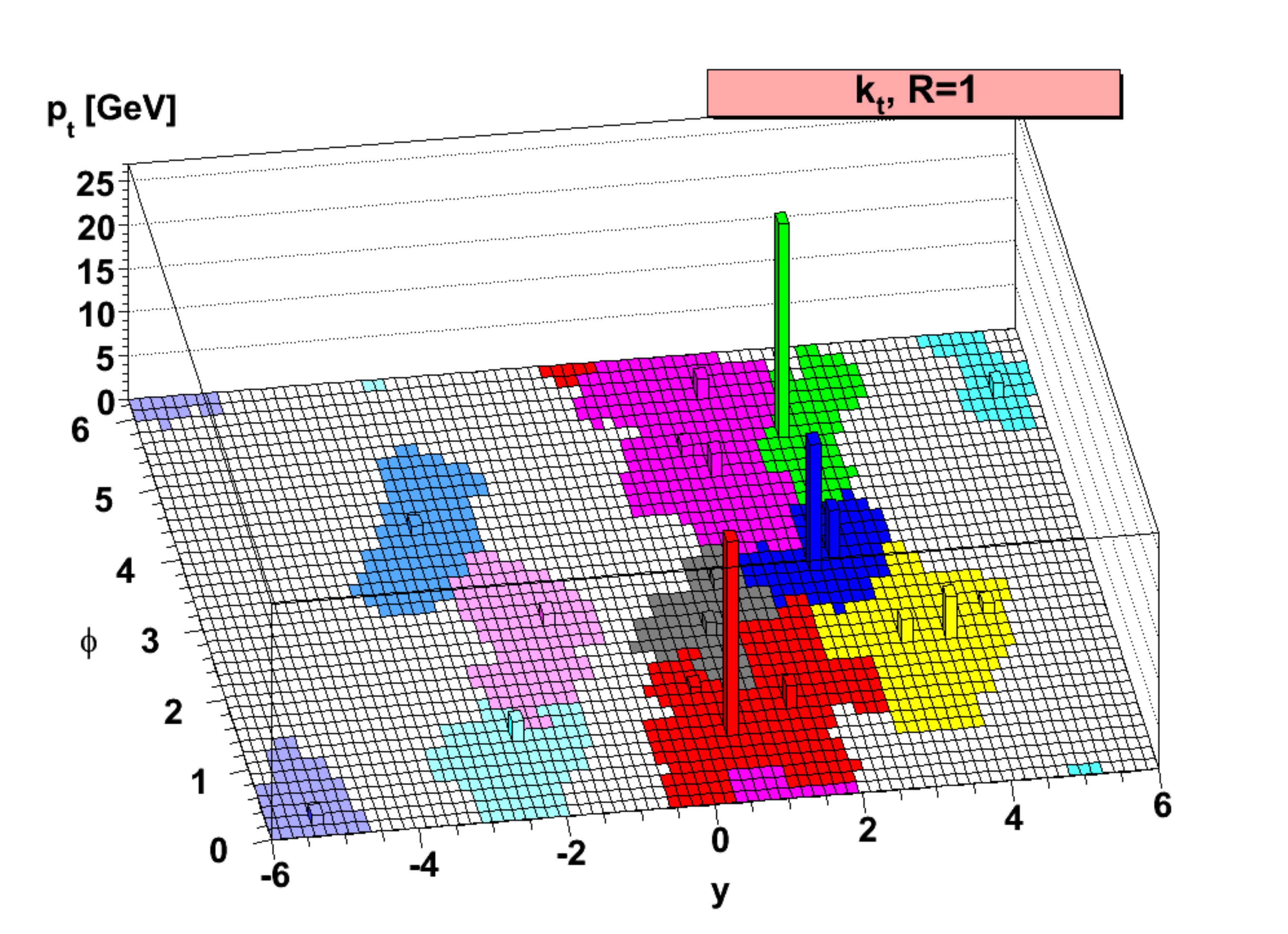}
    \includegraphics[scale=0.2]{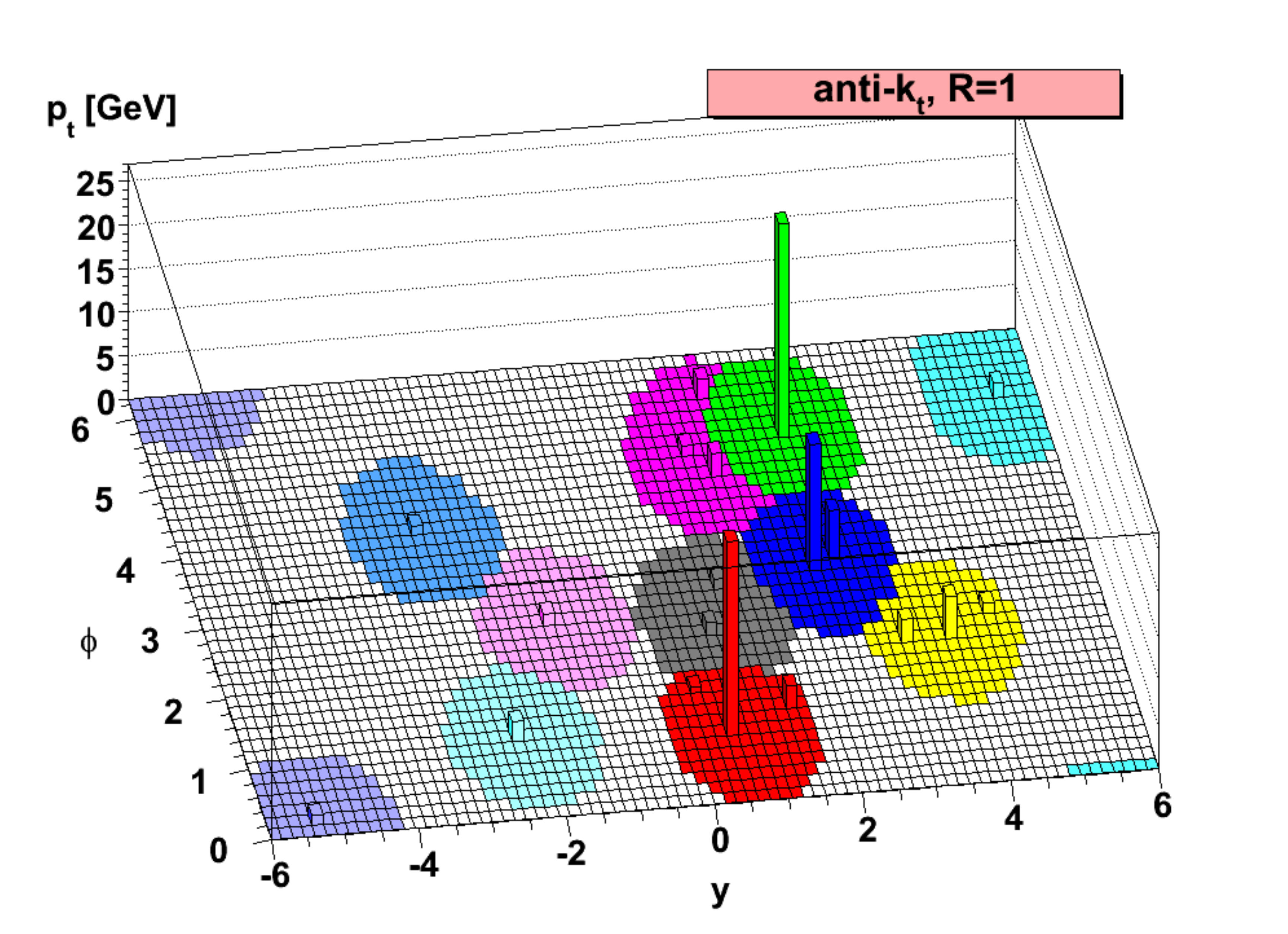}
  }
  \caption{Comparison of jet catchment areas for two different jet
    clustering algorithms\cite{Cacciari:2008gp}.  All particles within
    a jet's catchment area will be clustered into the
    jet.\label{fig:jetalgs}}
\end{figure}

Cacciari {\em et al.} pointed out recently that the case $p=-1$ also
defines a useful algorithm, the anti-kT
algorithm~\cite{Cacciari:2008gp}.  This choice means that the
particles around the hardest particle are clustered first.  The
anti-kT algorithm guarantees a cone-like geometry with well-defined
jet borders around the highest momentum particles (see
Fig.~\ref{fig:jetalgs}), but it maintains the infrared safety and
collinear safety of the sequential recombination family.  This
algorithm has become a preferred jet algorithm for LHC experiments,
along with a modified stable infrared-safe cone algorithm called
SISCone~\cite{Salam:2007xv}.

Total energy measured for a jet at the detector level must be
corrected to match the energy at the particle level, and the
calibration of the jet energy scale is a major experimental
uncertainty in signatures with hadronic jets~\cite{Bhatti:2005ai}.  In
short, the following effects are included in the calibration: varying
detector response due to non-linearities or uninstrumented regions,
mixed electromagnetic and hadronic showers in the same calorimeter,
overall absolute energy scale calibration (assuming differences in
relative response have been treated), and loss or gain of particles in
the region defined by the jet area.  These effects are estimated using
calibration data samples in the jet-jet or $\gamma$+jet signatures,
where the true jet energy can be estimated from the other object's
recoil.  Because there are no sufficiently large data samples of jets
at the highest energies, jet calibration at those energies is based on
extrapolation or on Monte Carlo simulation.

\begin{figure}
  \centerline{\includegraphics[scale=0.45]{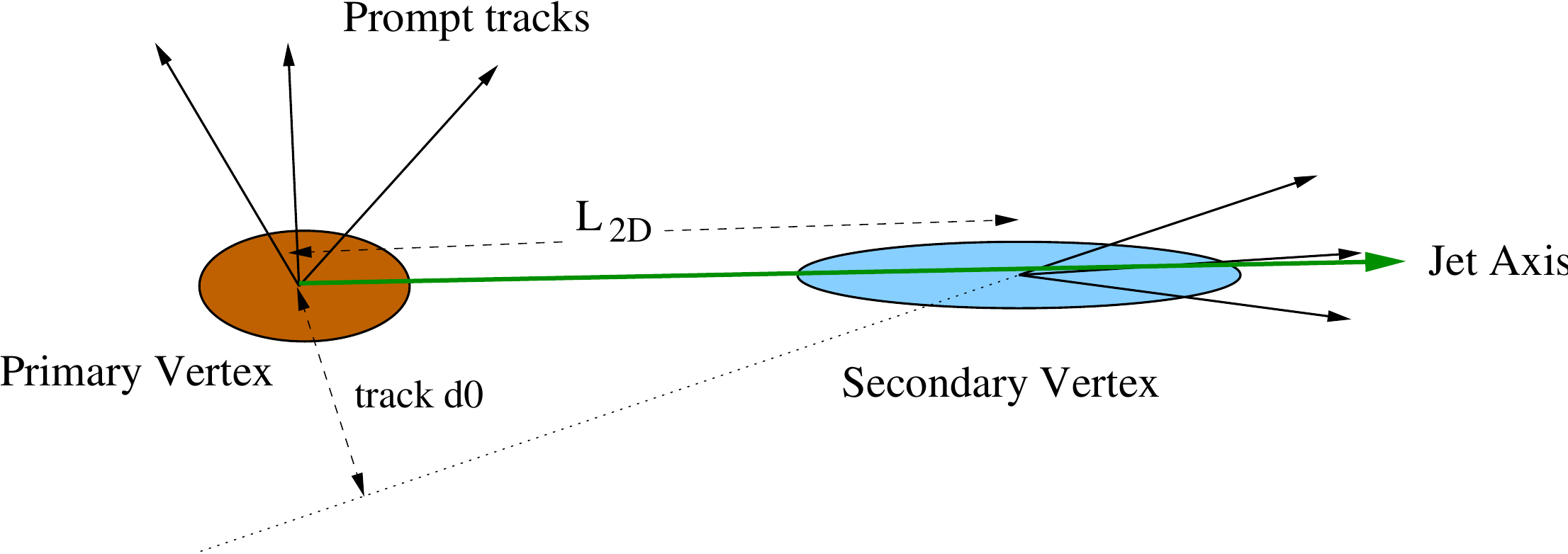}}
  \caption{Displaced secondary vertex from decay of a long-lived
    particle.  Tracks from the decay are not expected to point back to
    the primary vertex as prompt tracks do.  Flavor tagging
    (b-tagging) algorithms are designed to identify tracks with
    significant impact parameter $d_0$ and a vertex with significant
    decay length $L_{2D}$.\label{fig:vertexing}}
\end{figure}

Certain jets, specifically those associated with heavy quarks ($b,c$),
have several special properties due to the quarks themselves.  The
heavy quarks stand out for their long lifetimes (due to CKM
suppression), large mass with respect to their decay products, and
high multiplicity decays.  These properties give rise to a distinct
decay geometry, shown in Fig.~\ref{fig:vertexing}.  The momentum
vectors of decay products from the $B$ point to a secondary vertex,
not the primary interaction vertex, and the distance between the two
vertices depends on the $b$ lifetime, if we use the spectator model
approximation for the decay of the heavy quark hadron.  Typical values
before boost factors are $c\tau_b=500\,\mu\mathrm{m}$,
$c\tau_c=500\,\mu\mathrm{m}$, and $c\tau_\tau=90\,\mu\mathrm{m}$.
Because finding a common vertex for two or more tracks is one of the
most challenging problems in tracking, the impact parameter of a track
is used as a proxy to determine if it is consistent with having come
from the primary interaction vertex.  If a large number of tracks in a
jet are inconsistent with the primary vertex, then it is likely that
there is a heavy flavor hadron in the jet.  One of the primary
motivations for developing precision solid-state tracking detectors,
which are even sometimes named ``vertex detectors,'' is to measure
track parameters precisely enough to allow for vertexing.  The same
parameters are also use for exact impact parameter measurements with
10\% precision on values of $\mathcal{O}(300\,\mu\mathrm{m})$.

\begin{figure}
  \centerline{
    \includegraphics[scale=0.3]{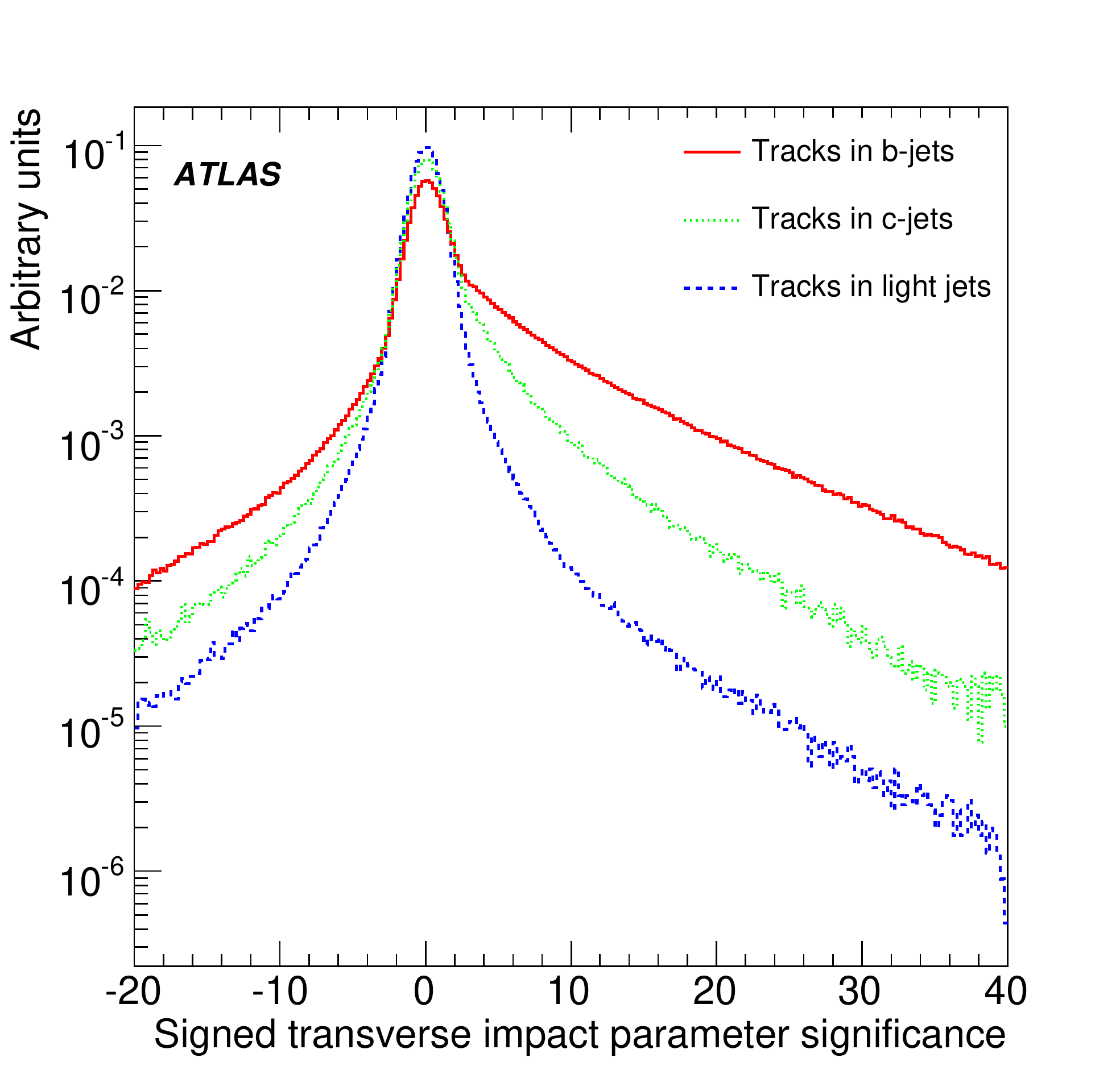}
    \includegraphics[scale=0.3]{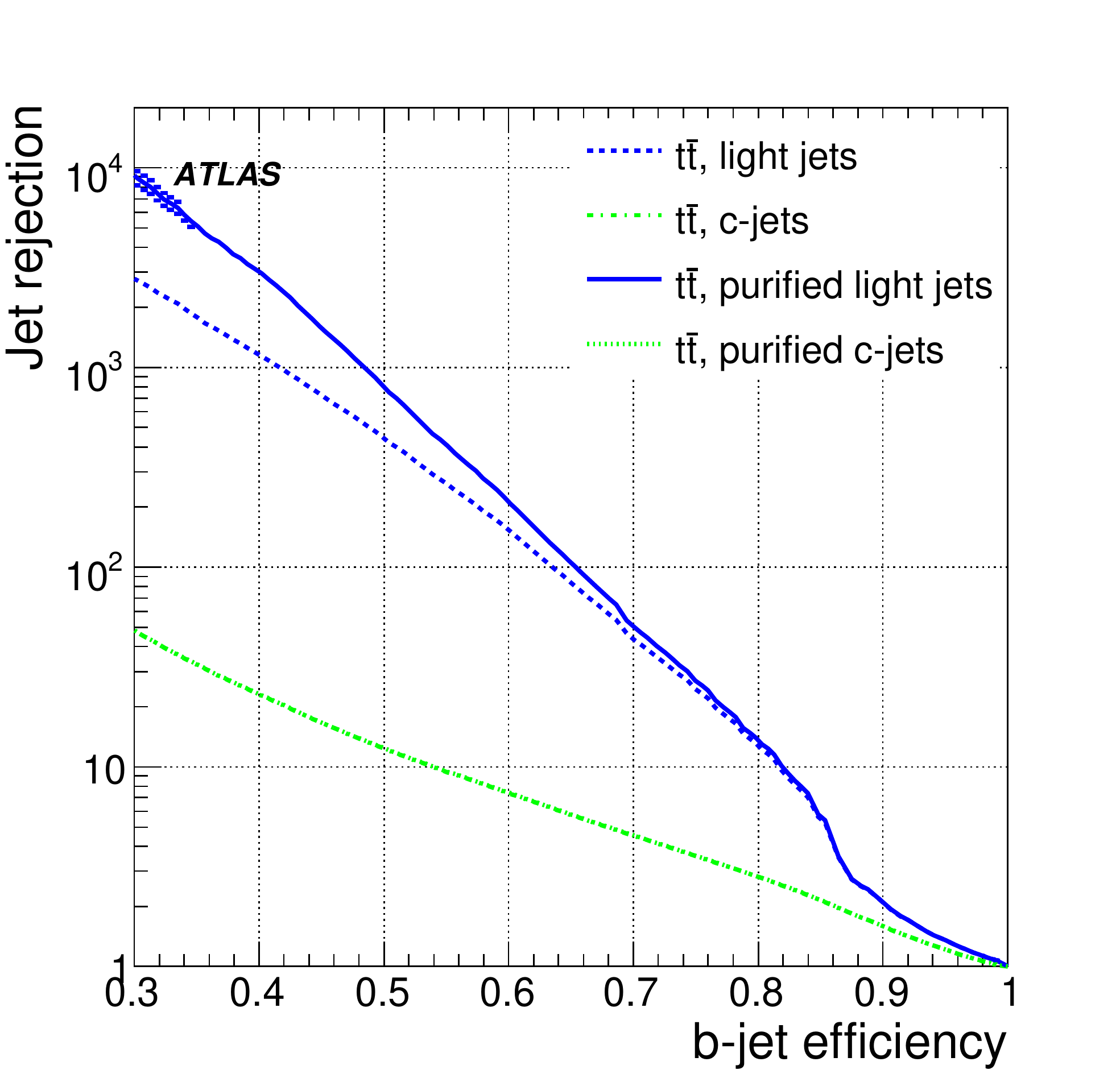}
  }
  \caption{Impact parameter significance distribution (left) and
    flavor tagging peformance (right) from the ATLAS
    experiment~\cite{Aad:2009wy}.  Flavor tagging algorithms exploit
    the difference in the impact parameter distributions to select
    jets from bottom quarks while rejecting jets from light quarks
    ($u,d,s$) and, to a lesser extent, jets from charm
    quarks.\label{fig:btagperf}}
\end{figure}

Flavor tagging algorithms are not limited to decay lengths and impact
parameters.  Identifying medium-$p_T$ leptons from semileptonic heavy
flavor decays provides an independent tagging mechanism.  Since charm
quarks also have long lifetimes and semileptonic decays, we use
discriminating variables based on the mass of all particles in the
secondary vertex to distinguish $b$-jets from $c$-jets.  Ultimately,
multivariate techniques combine information from all of these
measurements to give powerful separation between $b$, $c$ and light
quark jets.  Tighter requirements on $b$-jets reduce $c$ and light
quark jet contamination but also reduce the $b$-tagging efficiency,
as shown in Fig.~\ref{fig:btagperf}.

In the Standard Model, the proton-proton collision at a fixed
center-of-mass energy is in fact a parton-parton collision between
partons of unknown energy.  As a result, the longitudinal momentum of
the initial state is completely unknown, and this complicates the
final state reconstruction.  The key is to realize that the initial
transverse momentum is well known; it is essentially 0 because of the
small horizontal emittance of the beams and the low energy scale of
$\Lambda_\mathrm{QCD}$.  The final state transverse momentum is
expected therefore to also equal 0, and any deviation can be
interpreted as missing transverse momentum or ``missing transverse energy''
($E_T^\mathrm{miss}$ or MET),
presumably due to non-interacting particles produced in the
interaction but not detected.

The LHC experiments define the missing transverse energy as the
opposite of the vector transverse sum of all detected particles.  Such
a measurement is only relevant if the detectors are nearly hermetic to
both charged and neutral particles; this has put strict requirements
on the hermeticity of the experiments.  Corrections are applied for
the detector response to muons and jets, and in some cases information
from tracking and calorimetry is combined to optimize the missing
energy reconstruction.  The missing energy scale and resolution are
calibrated using events known to have specific missing energy, e.g.,
$Z(\rightarrow \nu\bar\nu)$+jets events.  It is important to calibrate
several different points to enable extrapolation to the higher values
of $E_T^\mathrm{miss}$ we expect in new physics signatures.  Like all
energy measurements, the absolute resolution on the missing transverse
energy scales as $\sqrt{E}$; early measurements of the ATLAS
resolution yielded $\sigma=0.57\sqrt{E\,(\mathrm{GeV})}$.

\begin{quote}
{\em Exercise 5:} The cone algorithm for jet clustering has its
drawbacks, but it does have one redeeming quality.  Use the definition
of the pseudorapidity $\eta$ to show that a cone size of $\Delta R
\equiv \sqrt{(\Delta \eta)^2 + (\Delta \phi)^2}$ is invariant under
boosts along $\hat{z}$.
\end{quote}

\begin{quote}
{\em Exercise 6:} Use Fig.~\ref{fig:vertexing} to show that, for a
typical track from a $B$ hadron decay, $d_0 \sim (c\tau_B)$, assuming
the kick transverse to the jet axis is due to the large mass of the
$B$ hadron.  Hint: consider the angle $\theta$ the track makes with
respect to the $b$-jet axis.  (This relation shows that the $B$ mass
is not as important as the lifetime when we consider track impact
parameters, and it explains why tracks from charm decay have $d_0$
values of similar magnitude.)
\end{quote}

\section{Searches for Higgs Bosons\label{sec:higgs}}

The search for the Higgs boson, whether in the Standard Model or
beyond, is a key goal for understanding the physics of the terascale.
The $W$ and $Z$ boson and $t$ quark masses near the
$100\,\mathrm{GeV}$ scale give quantitative constraints on the
Standard Model Higgs boson mass, the only unknown parameter in the
electroweak sector.  In particular, precision measurements of
$m_W,m_Z,m_t$ constrain Higgs loop contributions and favor low Higgs
masses, below $200\,\mathrm{GeV}$.  Direct experimental searches at
LEP and Tevatron rule out $m_H<114\,\mathrm{GeV}$ and
$158<m_H<175\,\mathrm{GeV}$, respectively~\cite{Barate:2003sz,:2010ar}.

\begin{figure}
  \centerline{
    \includegraphics[scale=0.36]{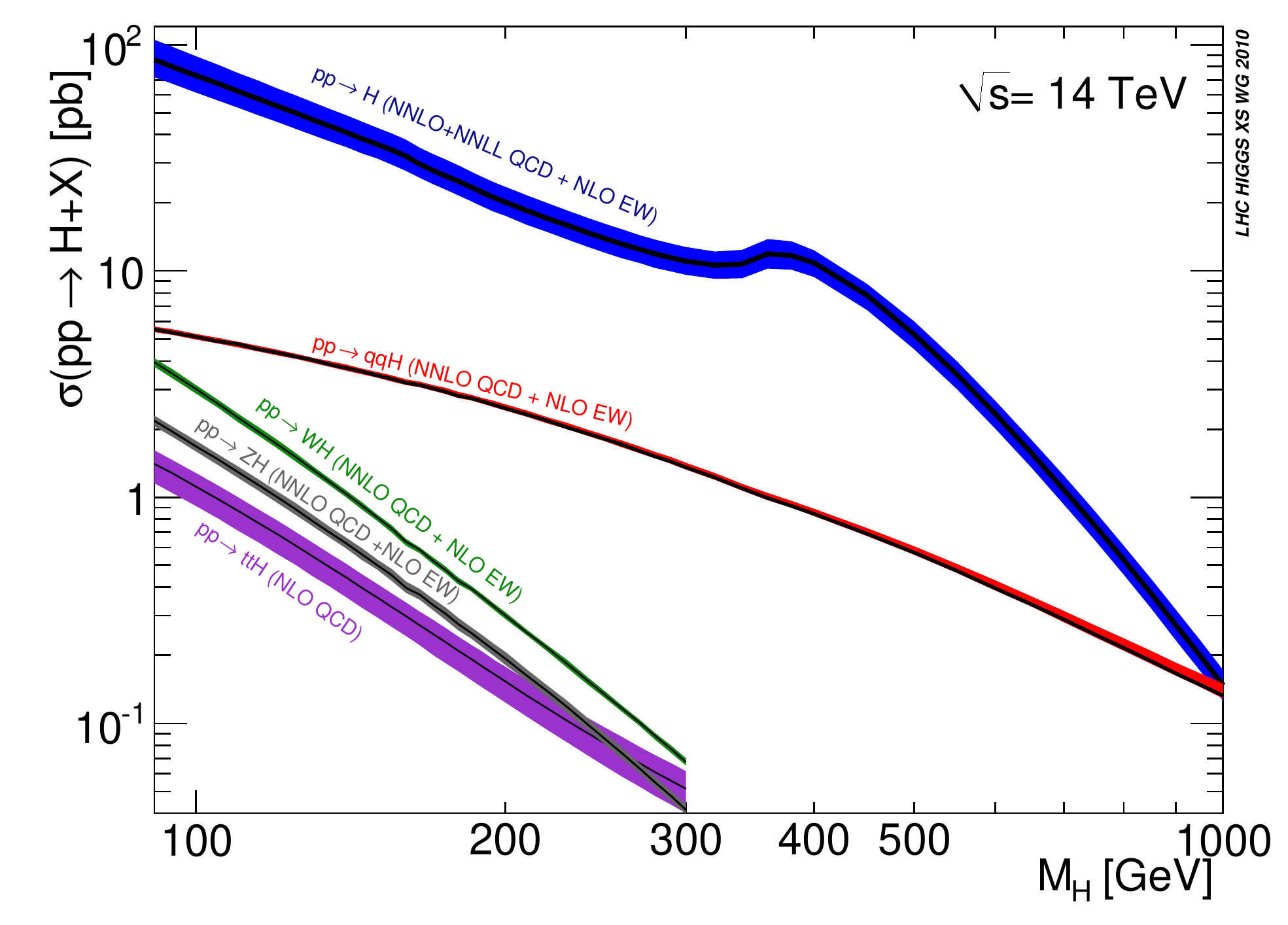}
    \includegraphics[scale=0.27]{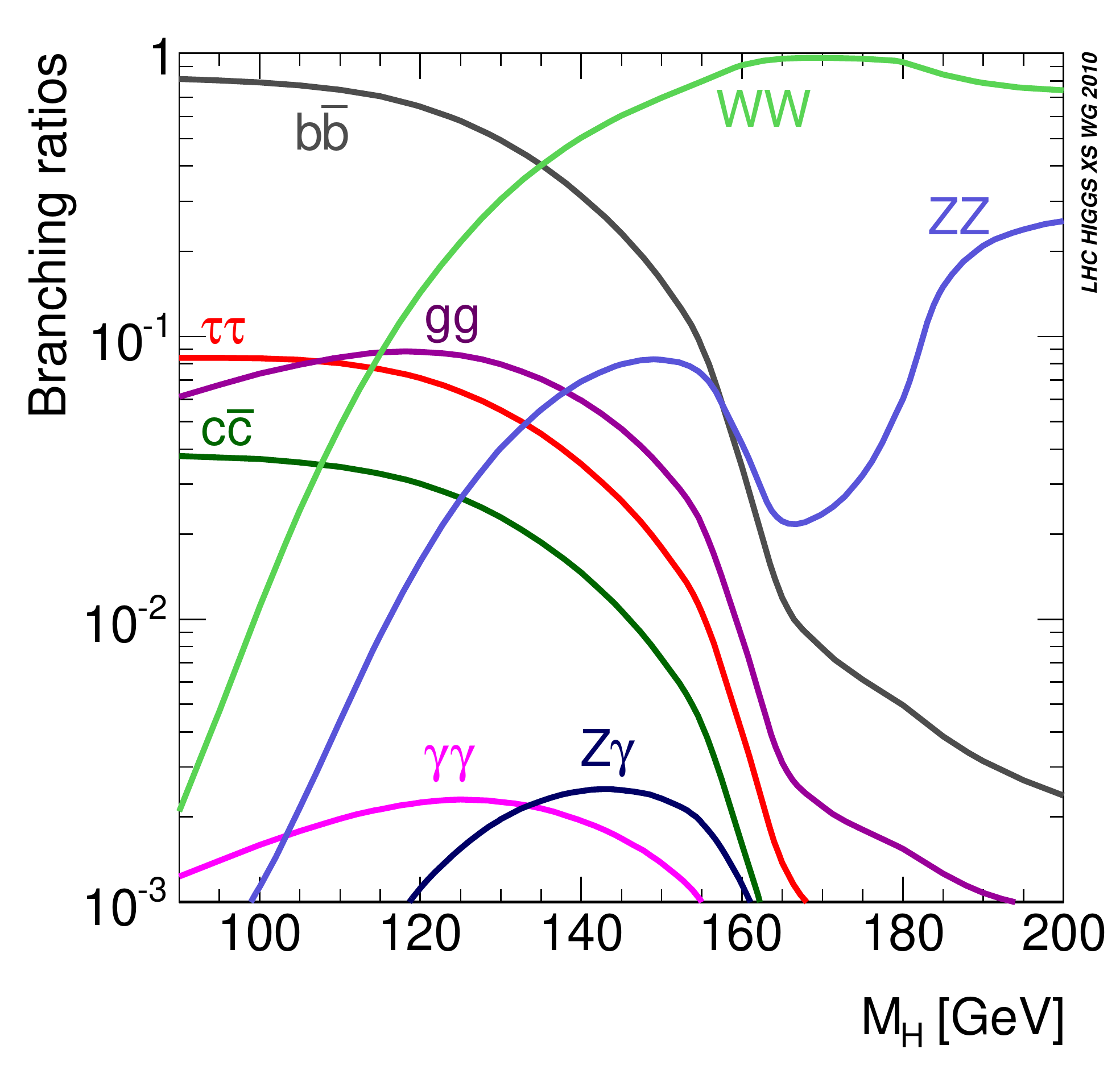}
  }
  \caption{Calculations of the Standard Model Higgs boson production
    $pp$ cross section (left) and branching fractions
    (right)~\cite{LHCHiggsCrossSectionWorkingGroup:2011ti}.\label{fig:higgsxsecbr}}
\end{figure}

Standard Model Higgs boson production cross sections and branching
ratios depend only on the Higgs mass.  As seen in
Fig.~\ref{fig:higgsxsecbr}, the gluon fusion production mechanism
dominates, but other production mechanisms are important for
signatures with small background contributions.  The branching
fractions change quickly with increasing $m_H$ as phase space for new decay
channels opens.  One notable feature is the dominant $WW$ decay even
above $2m_Z$; this is explained by the Standard Model Lagrangian term
\begin{equation}
\mathcal{L} \sim (2 M_W^2 H W_\mu^+W^{-\mu} + M_Z^2 H Z_\mu Z^\mu) .
\end{equation}

A successful Higgs boson search would show not only a discrepancy in the
data with respect to the Standard Model prediction (without Higgs),
but also consistency with expected Higgs production.  We choose a
specific experimental signature or ``channel'' and develop an event
selection to reject backgrounds from SM physics processes.  The data
sample may include contributions from background processes and
putative signal events, but the goal is to measure the background
contributions directly from data, if possible, to avoid any bias or
errors in simulation.  

A vital, if controversial, part of the Higgs search is the statistical
interpretation of observed results in the context of a Higgs boson
production hypothesis.  What is the probability that the observed
dataset is consistent with background-only production?  With signal
plus background production?  Suppose 40 background (SM) events are
expected in a search for a model that predicts 10 events from a new
physics signal.  Can the new physics model be excluded definitively if
40 events are observed?  Can the Standard Model be excluded if 50
events, or even 60 events, are observed?  

Much work in the past decade has focused on bringing sophisticated
statistical tools to bear on this question in particle physics.  The
most common shorthand for presenting results is a re-interpretation in
terms of a Gaussian distribution.  If the observed experiment is
considered as one of many possible experiments, given a certain model,
then it is possible to calculate where the observed experiment lies in
the distribution of the ``pseudo-experiments'' and convert its
percentile to a number of ``sigma.''  That is, if $\alpha$ is the
probability to measure a less likely value than the observed
experiment, then $\alpha=0.3173$ corresponds to a $1\sigma$ deviation.
(It is important to be aware of the distinction, sometimes overlooked,
between one-sided and two-sided definitions of
$\alpha$~\cite{Nakamura:2010zzi}.  Two-sided definitions are typical
for measurements, while one-sided is often used for counting events
when the signal is unknown.)  

Usually we are interested in testing two complementary hypotheses,
that of background-only production ($s=0$) and that of
signal+background production ($s>0$).  If the data favor the latter
hypothesis and strongly disfavor the former, then we have a discovery.
Experiments often report results in terms of a likelihood ratio
$LR=L_{s+b}/L_b$ and by asking the following: how often can certain
values of the LR be expected from an experiment in the presence of
signal?  For a discovery we talk about excluding the background
hypothesis at $>5\sigma$, which is $P_b<10^{-7}$.

Most of the progress in the past five years has been on the treatment
of systematic uncertainties, which reflect the inherent uncertainty in
the number of background and signal expected in the datasets.  Poorly
constrained backgrounds can doom a Higgs search just as surely as low
integrated luminosity.  An overview of some methods, with technical
results somewhat beyond the scope of these lectures, is given in
Ref.~\cite{Gross2008}.

Now it is instructive to introduce four of the main Standard
Model Higgs boson searches at the LHC and to have a peek at one that
may be important in the future.

For low mass Higgs bosons ($m_H<130\,\mathrm{GeV}$), one might expect
to search for Higgs resonances produced in gluon fusion and decaying
to $b\bar b$ pairs.  Unfortunately, non-resonant $gg\rightarrow b\bar
b$ production has an enormous production rate at the LHC, about 6
orders of magnitude greater than Higgs production!  Fortunately, there
are two other possibilities.

Higgs decays to tau lepton pairs (approximately 10\% branching
fraction) do not suffer from the $gg$ background because the tau is
not colored.  The dominant background process is $Z\rightarrow
\tau^+\tau^-$, but the $Z$ mass resonance is well below the search
region.  After two tau lepton candidates have been reconstructed in an
event, using algorithms that identify fully leptonic decay or hadronic
decay, the missing transverse energy is used to estimate the energy
taken by two or more neutrinos.  With no other information, the
kinematic system would be underconstrained, and mass reconstruction of
the $\tau\tau$ system would be impossible.  If one assumes the tau
leptons from Higgs decays are highly boosted, then the neutrinos in
the tau decay all have the same momentum as their sister decay
products.  With this trick, it is possible to reconstruct the full tau
momentum and calculate the invariant mass of the Higgs candidate (see
Exercise 8).

\begin{figure}
  \centerline{
    \includegraphics[scale=0.25]{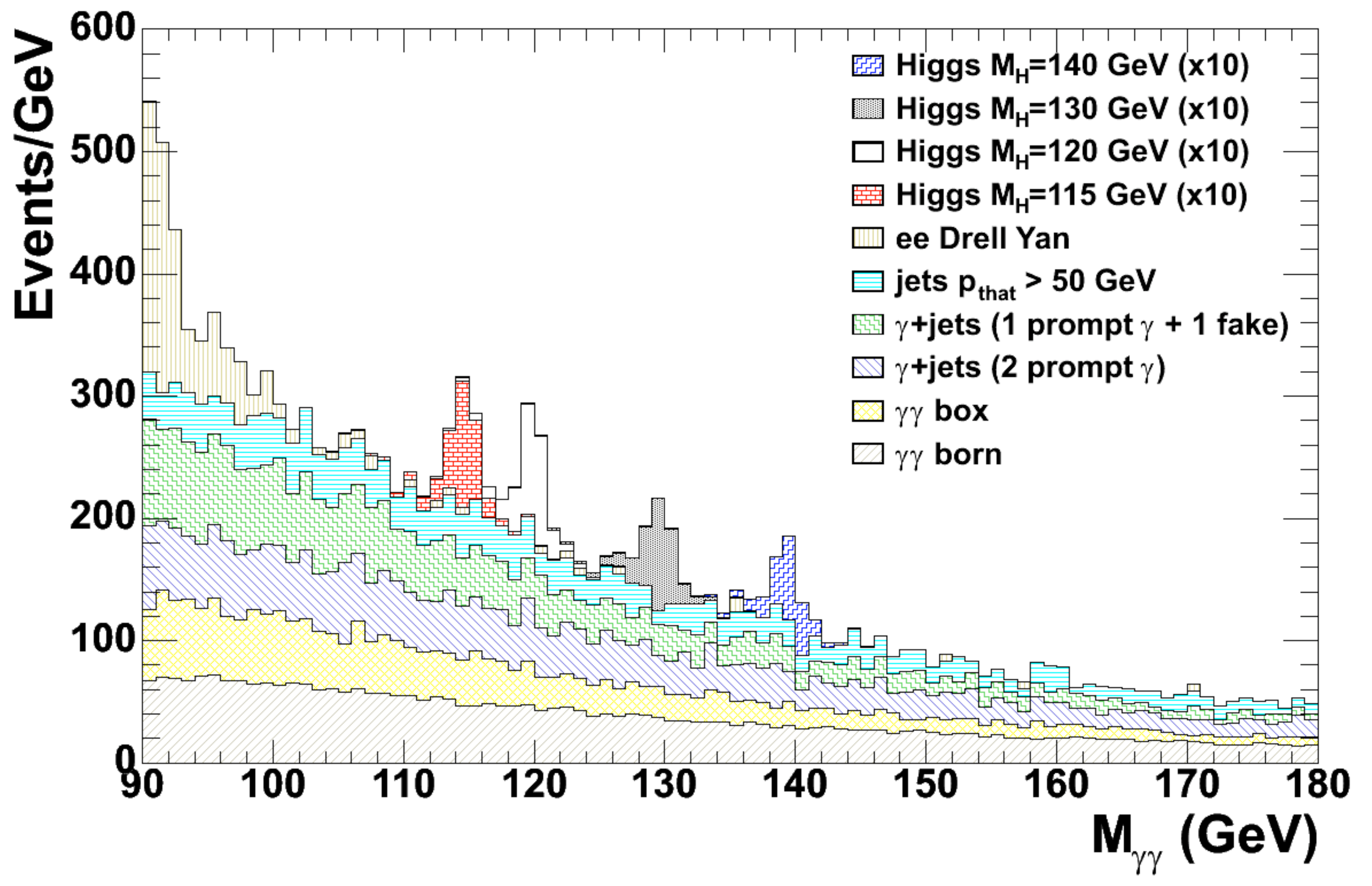}
  }
  \caption{Invariant mass spectrum of diphoton candidates selected in
    the low-mass Higgs boson search~\cite{0954-3899-34-6-S01}.  The
    significance of the diphoton resonances depends on the mass
    resolution of the experiment and the level of background
    underneath the signal peak.\label{fig:cmsdiphoton}}
\end{figure}

Higgs decays to two photons are a tiny fraction of all decays, but
there are no diphoton resonances in the Standard Model above
$1\,\mathrm{GeV}$, so any sign of a resonance would be a clear
indication of new physics.  The LHC experiments, particularly ATLAS
and CMS, have been designed to have excellent resolution for both
photon energy and direction, and significant effort has gone into
rejecting fake photons (misidentified electrons or jets) to reduce
background.  The non-resonant diphoton background can be measured as a
falling distribution in data, as shown in Fig.~\ref{fig:cmsdiphoton}
compared to the signal expectations for Standard Model Higgs
production.  This search is limited only by the number of events that
can be collected, given the small branching ratio.

\begin{figure}
  \centerline{
    \includegraphics[scale=0.3]{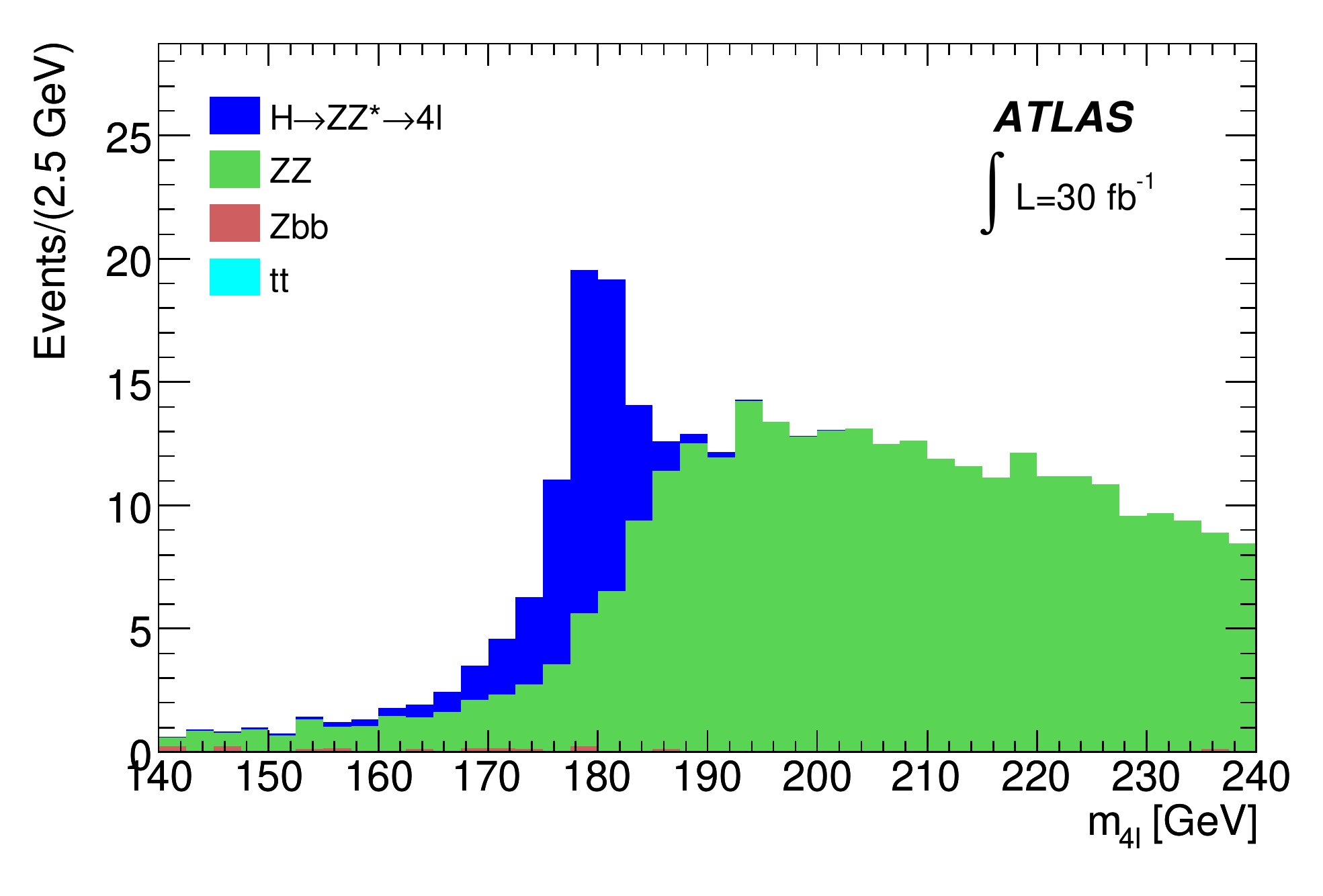}
    \includegraphics[scale=0.3]{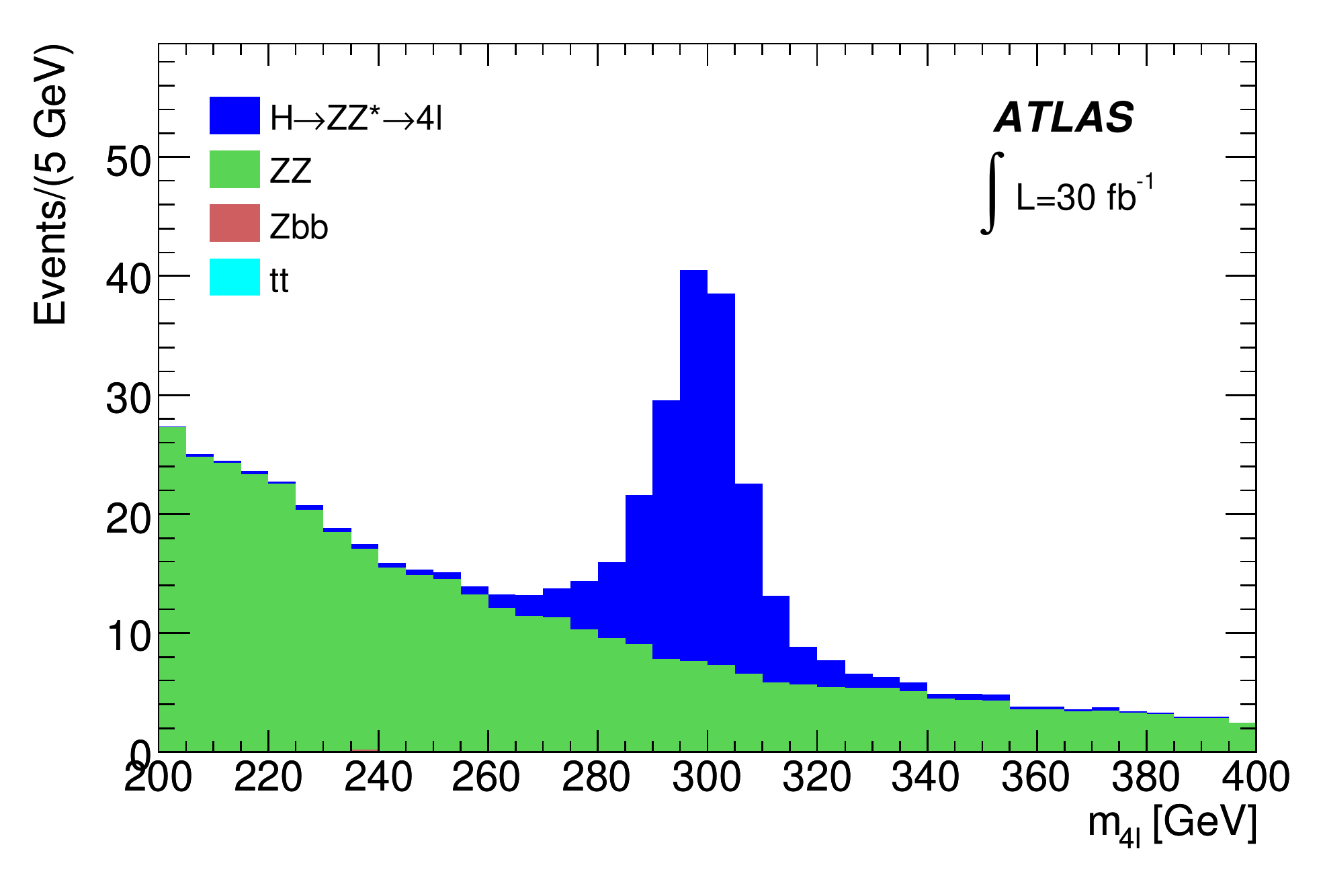}
  }
  \caption{Reconstructed 4-lepton mass for signal and background
    processes, assuming $m_H=180\,\mathrm{GeV}$ (left) and
    $m_H=300\,\mathrm{GeV}$
    (right)~\cite{Aad:2009wy}.\label{fig:hzz4l}}
\end{figure}

For high mass Higgs bosons ($m_H>170\,\mathrm{GeV}$), the
$ZZ\rightarrow \ell^+\ell^-\ell^+\ell^-$ decay channel offers another
clean signature.  Even though there is non-resonant $ZZ$ Standard
Model production, the reconstructed Higgs resonance would stand out
clearly, as shown in Fig.~\ref{fig:hzz4l}.  Again, the LHC experiments
are designed to make precise measurements of the high-$p_T$ electrons
and muons, even for large Higgs masses; this defines the target
resolution for the muon spectrometers.

For medium mass Higgs bosons ($130<m_H<170\,\mathrm{GeV}$), the
dominant decay is $H\rightarrow W^+ W^-$.  This signal process has a
large rate using all production mechanisms, and the decay to a dilepton
signature is clean.  There are two challenges for searches in this
channel.  First, the presence of two high-$p_T$ neutrinos means there
is no invariant mass peak for the reconstructed Higgs.  (Since the $W$
bosons are not boosted, the reconstruction trick from the tau channel
cannot be re-used.)  Second, direct $WW$ production and top quark
pairs have similar dilepton+$E_T^\mathrm{miss}$ signatures, but even
here there is one extra trick for selecting $WW$ pairs from Higgs
decay.  Because the Higgs is a scalar boson with spin 0, the two $W$
bosons from Higgs decay must have opposite spin, and the leptons from
$W$ decay tend to be closer in direction than in the $t\bar t$ case.
The angle between leptons is one of several input variables for
multivariate tools that separate Higgs signal from SM background.
This technique has already been used at the Tevatron to exclude
certain Higgs mass hypotheses between 158 and 175\,GeV~\cite{:2010ar}.

\begin{figure}
  \centerline{ \includegraphics[scale=0.5]{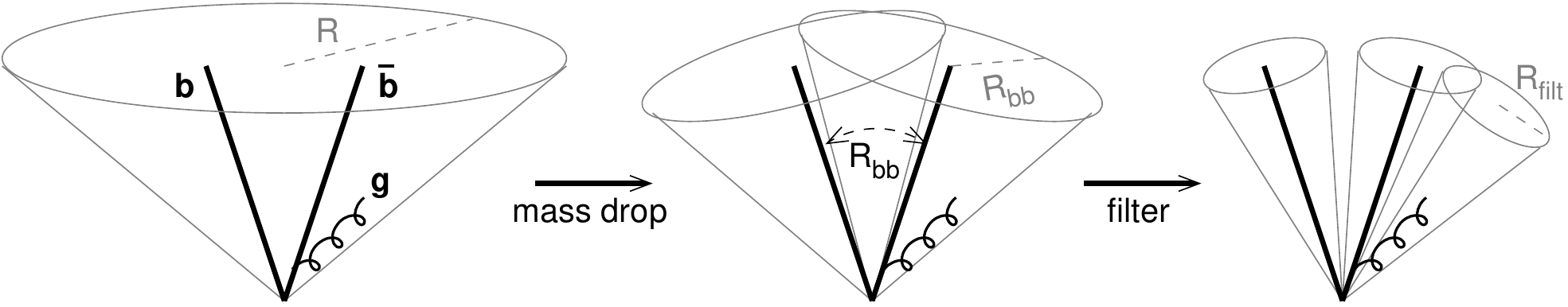} }
  \caption{Description of jet algorithm to identify boosted jets and
    substructure within~\cite{Butterworth:2008iy}.  The algorithm is
    designed to identify cleanly two $b$-jets from $H\rightarrow b\bar
    b$ decay, but it can be used for any decay of a particle to
    products having much lower mass.\label{fig:boostedjets}}
\end{figure}

The peek into the future regards the dominant $b\bar b$ decay channel
for low-mass Higgs bosons.  Since electroweak fits favor low Higgs
masses, this decay channel would seem to be of prime importance in the
Higgs search, but the $b\bar b$ jet background limits its sensitivity.
(If the $b\bar b$ mass resolution were as good as the $\gamma\gamma$,
a peak might still be resolved, but the jet energy resolution is
limited by fluctuations in the hadronic shower.)  One way to reduce
the $b\bar b$ background is to require associated production of $W,Z$
or $t\bar t$, and all of these associated production channels are
being studied.  A novel idea focuses the $WH/ZH$ search to the region
of phase space where the vector boson has large transverse momentum.
Events in this region have a highly-boosted Higgs boson and $b\bar b$
decay products observed in a single fat jet.  By shrinking the jet
clustering radius until the two $b$ subjets are resolved (as in
Fig.~\ref{fig:boostedjets}), it is possible to compare the masses of
the subjets to the mass of the boosted Higgs
jet\cite{Butterworth:2008iy}.  The large mass drop from the parent
Higgs jet to both daughter $b$ subjets is nearly unique; only the $WZ$
decay to $b\bar b$ has a similar reduction.  The background from
two-body $gg\rightarrow b\bar b$ decay is greatly reduced, and the
dominant background becomes $W/Z$+light jet production.  The benchmark
calibration for this new technique is the reconstruction of the $Z$
peak in Standard Model $WZ$ and $ZZ$ production.

Because the Higgs production cross section is small, several promising
channels must be combined in the low-mass region to ensure
sensitivity.  When added to the powerful high-mass channels ($WW,ZZ$),
these searches guarantee the LHC experiments will have something
definitive to say about the Standard Model Higgs boson in the near
future, perhaps with several $\mathrm{fb}^{-1}$ of 7\,TeV data.

\begin{quote}
{\em Exercise 7:} There is a quick way to estimate the ``number of
sigma'' significance of a signal observed above the background
expectation.  Use Poisson probabilities to define a likelihood ratio
$L = P_{s+b}/P_b$ and show that to good approximation
\[
\sigma \equiv \sqrt{2\ln L} =
\sqrt{2\left[\left(s+b\right)\ln\left(1+\frac{s}{b}\right)-s\right]},
\]
where $s$ and $b$ are the expected numbers of signal and background
events, respectively.
\end{quote}

\begin{quote}
{\em Exercise 8:} (suggested by M. Strassler): Consider an event $pp
\rightarrow Z + \gamma$, where the $Z$ decays to $\tau^+\tau^-$. Use
the fact that, although the neutrinos carry off energy, they do not
significantly alter the directions of the tau leptons' other decay
products, since the taus are highly boosted. Show that the $Z$ boson
mass can be reconstructed using only the photon momentum, the observed
missing $p_T$ , and the tau momentum directions (not their energies).
This technique can also be used to measure the Higgs boson mass in
$H\rightarrow \tau^+\tau^-$ decays.
\end{quote}

\section{Searches for Physics beyond the Standard Model\label{sec:bsm}}

Searches for physics beyond the Standard Model (BSM) follow one of two
approaches.  The search strategy may focus on a specific model, or it
may target any discrepancy from the Standard Model expectation.  Both
of the strategies are used in the LHC experiments, which hunt for
general features in data that may correspond to a wide range of BSM
models.  Results in this section are a sampling of techniques used in
new physics searches.

Perhaps the most straightforward signature shared by BSM models is the
total event energy.  New physics related to the terascale often has
energies near this scale.  The outgoing events in the hard scatter set
the event scale, which is near the mass of the heavy new particles.
Whatever the exact nature of their subsequent decay, the event energy
scale is roughly preserved.  As a result, the total event energy is a
good estimate for the mass of new particles produced in pairs.  To
make this connection, it is best to focus on a robust definition of
the total events energy.

There are at least three common calculations corresponding to total
event energy.  The first is the simple $\sum E_T$, for which all
calorimeter energy is summed.  This definition does not cover
non-interacting particles, such as neutrinos or weakly interacting
massive particles, nor does it account for extra calorimeter activity
due to pileup events.  The second is the oft-misunderstood $H_T$,
usually defined as the scalar sum of missing transverse energy and the
transverse energies of identified jets and leptons.  This definition
works well if there are one or two non-interacting particles, but it
still suffers from pileup contamination and depends on the
identification of the physics objects.  The third is the effective
mass $M_\mathrm{eff}$, usually defined as the scalar sum of transverse
energies of the four hardest identified jets and the missing
transverse energy.  This definition suppresses low-energy
contributions from pileup events, but it does not capture the leptonic
parts of the new particles decay chains.  All three of these
definitions are used in various channels by the LHC experiments, where
they show good discriminating power between Standard Model background
and the new physics signal.

Many arguments have been advanced for higher-mass versions of SM
particles, and some of these correspond to resonances (invariant mass
peaks) of simple objects, such as leptons or jets.  Examples include
Kaluza-Klein towers of particles confined in extra dimensions,
$Z'/W'$, and $t'$, a 4th-generation up-type quark.  Reconstruction of
these resonances is straightforward, if the 4-momenta of all decay
objects are known.  The sensitivity to resonances is limited by
background contamination and invariant mass resolution, both of which
are high priorities for the LHC experiments.

If some of the decay products from the new resonance are invisible
(non-interacting), a simple invariant mass calculation will not
capture the signal.  In these cases, the transverse mass
\begin{equation}
m_T^2 = 2E_{T1}E_{T2}\left[ 1- \cos(\Delta\phi)\right]
\end{equation}
can be used, using as $E_{T}$ the missing transverse energy.  If two
particles decay to invisible daughters, as in cascade decay chains to
lightest supersymmetric particles, it is still possible to apportion
correctly the missing transverse energy.  For example, in the
supersymmetric decay chain
\begin{equation}
pp \rightarrow X +\tilde\ell^+_R \tilde\ell^-_R \rightarrow X +
\ell^+\ell^- \chi_1^0 \chi_1^0
\end{equation}
the final state neutralinos both appear as missing transverse energy.
If the $\chi_1^0$ mass is known, then the mass equivalence of the
slepton mothers gives enough constraints to construct a kinematic
variable whose distribution endpoint gives the slepton
mass~\cite{Lester:1999tx}.

\begin{figure}
  \centerline{
    \includegraphics[scale=0.25]{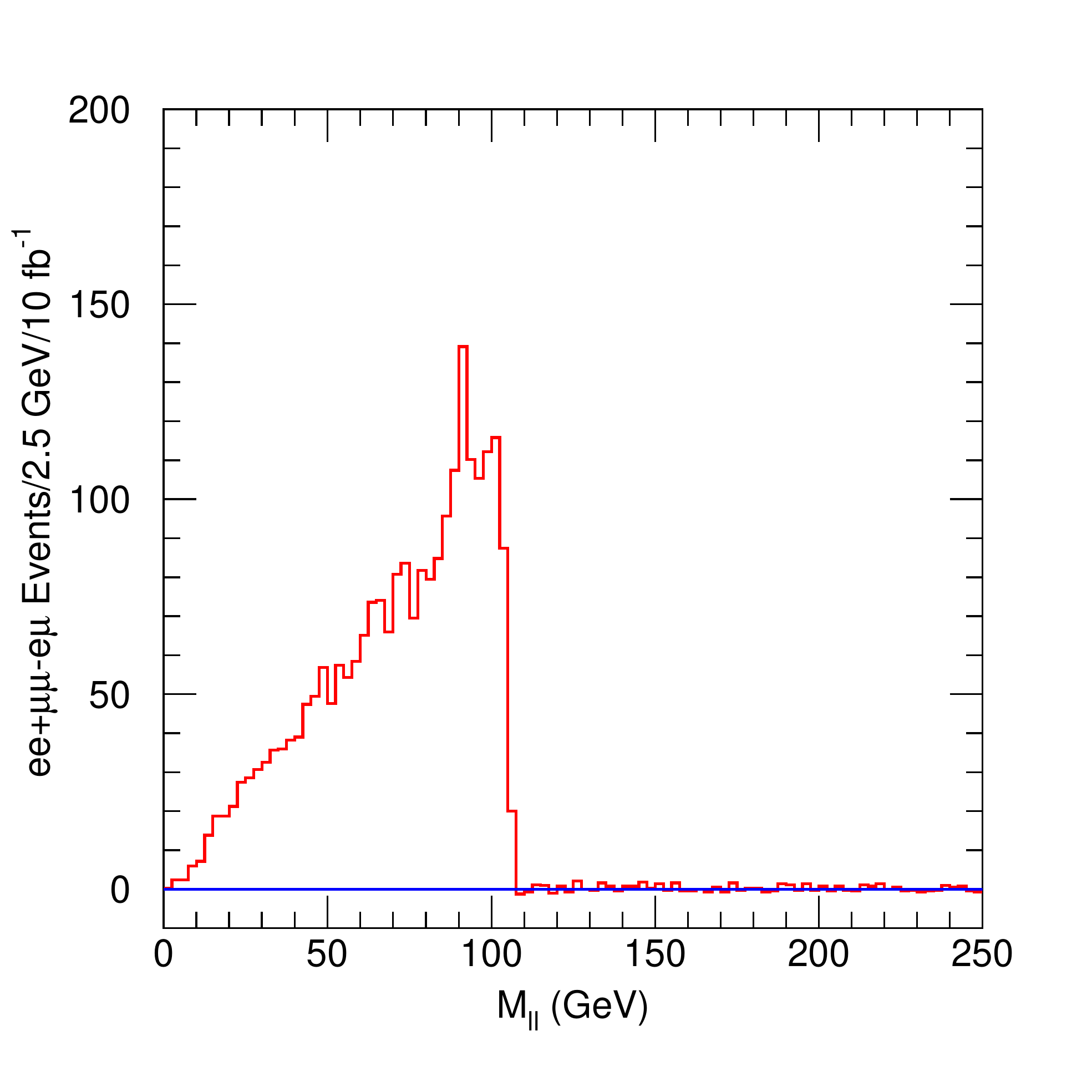}
    \includegraphics[scale=0.25]{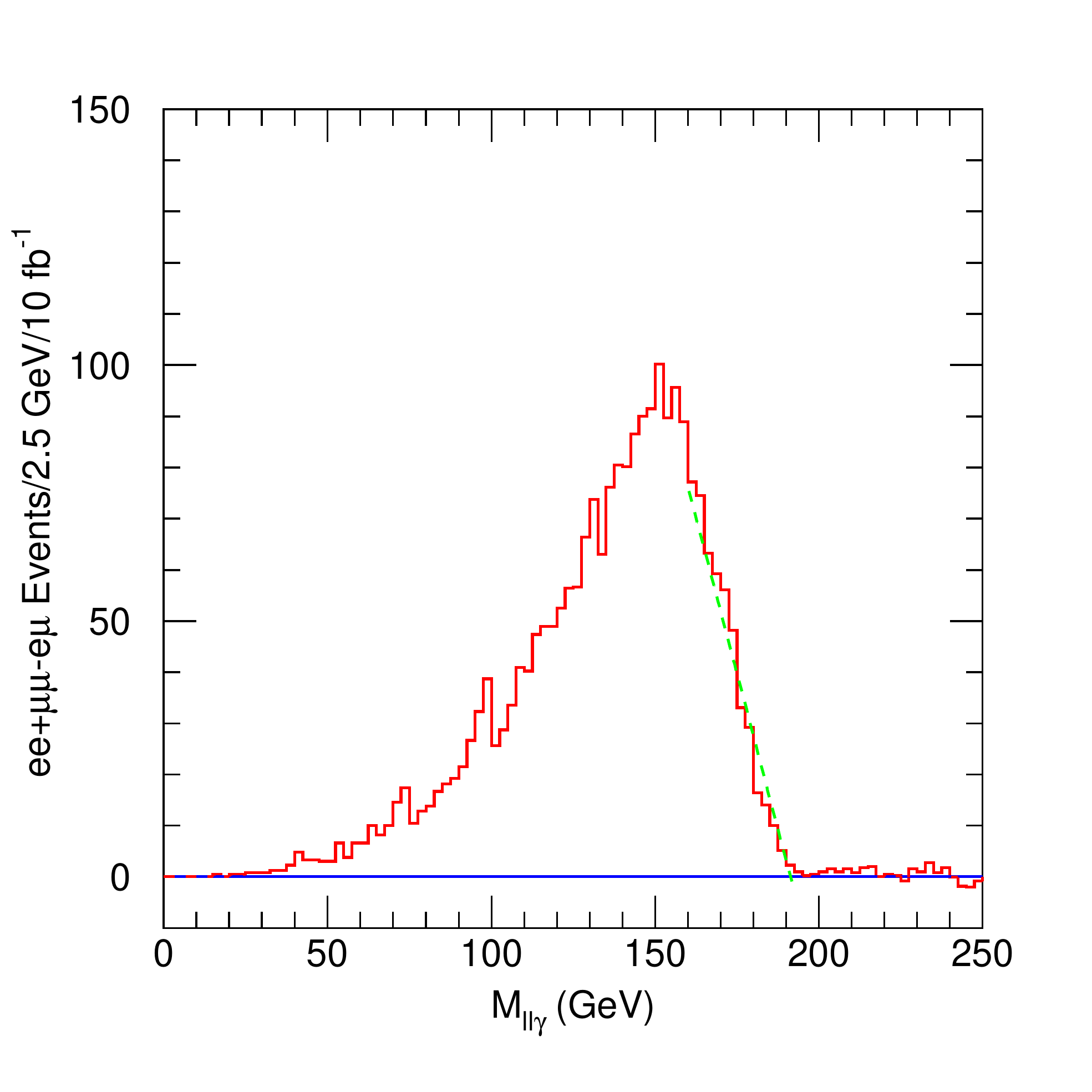}
  }
  \caption{Invariant dilepton mass distribution (left) and minimum
    dilepton+photon mass distribution (right) with background
    subtraction for GMSB signal events \cite{Hinchliffe:1998ys}.  The
    endpoints of the background-subtracted distributions can be used
    to determine the masses of intermediate particles in the new
    physics decay chain. These figures represent data corresponding to
    an integrated luminosity of
    $10\,\mathrm{fb}^{-1}$.\label{fig:gmsbdist}}
\end{figure}

For a more concrete example, consider the following decay in theories
of gauge-mediated supersymmetry breaking:
\begin{equation}
\tilde{\chi}_2^0 \rightarrow \tilde{\ell}^\pm \ell^\mp \rightarrow
\tilde{\chi_1^0} \ell^\mp \ell^\pm \rightarrow \tilde{G} \gamma
\ell^\pm \ell^\mp
\end{equation}
Assuming that the background in this dilepton channel can be estimated
using an opposite-flavor sample, the background-subtracted invariant
mass distribution (Fig.~\ref{fig:gmsbdist}) shows a sharp edge,
indicating a kinematic limit~\cite{Hinchliffe:1998ys}.  In this case,
the location of the edge is determined by the relation between the
neutralino and slepton masses.  The minimum mass for the
$\ell\ell\gamma$ system has a similar endpoint, given by the difference
of the neutralino masses.  By using these and other kinematic solution
endpoints, it is possible to reconstruct all of the masses in the
decay chain.

An alternative to targeted searches has emerged in the last decade.
Instead of developing event selections and kinematic variables
designed for each of many different models and decay chains, some
experiments have deployed general search strategies.  These programs,
count events in each of several high-$p_T$ object classes ($1\mu
1\mathrm{jet},1e2\mathrm{jet}$, etc.) and compare the results to SM
expectations~\cite{Abbott:2000fb, Aktas:2004pz, Aaltonen:2007dg}.  The
challenge is to describe completely the SM backgrounds for all
signatures at once!  Several discriminant distributions are considered
for each class, including the scalar $p_T$ sum of all objects, the
invariant mass (or transverse mass) of all objects, and the missing
transverse energy.  Any discrepancies between observed data and
expectations are flagged for further studies.

In conjunction with the rise in general searches, new emphasis has
been placed on simplified phenomenological models, which include the
gross effects of new particle mass spectra and decay chains without
focusing on the details of particle couplings and spin effects.  These
models have been successful in identifying experimental signatures
that may have been overlooked~\cite{Alves:2011wf}, and they offer hope
of helping match experimental observations with consistent theoretical
models of new physics~\cite{ArkaniHamed:2007fw}.

\section{Conclusion}

The Large Hadron Collider and associated experiments have been
designed and constructed to answer questions about physics at the
1\,TeV scale.  The size, scope, and details of the experiments stem
directly from the physics goals and requirements on measurements at
that energy scale.

The focus in the last few sections on Higgs boson searches and other
specific searches led naturally to a concentration on results from the
general-purpose ATLAS and CMS experiments, but the other LHC
experiments have been designed to pursue different physics goals that
are no less interesting.  All of the detector interactions, many of
the design considerations, and some of the analysis techniques are
being brought to bear on those goals as well.

A basic understanding of detector physics and practical limitations
makes interpretation of experimental results more exciting and
engaging, and the next few years may even bring news of exotic new
particles with unexpected signatures.

\section*{Acknowledgments}
My discussions with Matthew Strassler played an important role in the
development of this material, and I thank him for his suggestions.  I
also thank the TASI 2010 organizers -- Tom Banks, Michael Dine, and
Subir Sachdev -- for their efforts and patience, and the local hosts
K.T. Mahantappa, Tom DeGrand, and Susan Spika for their hospitality.

\providecommand{\href}[2]{#2}\begingroup\raggedright\endgroup

\end{document}